\documentclass{svjour3}                     
\smartqed  
\usepackage{graphicx}
\usepackage{braket}
\usepackage{orcidlink}
\usepackage{graphicx}
\usepackage{euscript,amssymb}
\usepackage{amsfonts}
\usepackage{amssymb}
\usepackage{amsbsy}
\usepackage{amsmath}
\usepackage{nccmath}   
\usepackage{color}
\usepackage{bbold}
\usepackage{hyperref}
\usepackage{physics}
\usepackage[caption=false]{subfig} 

\newcommand{\be}{\begin{equation}}
  \newcommand{\ee}{\end{equation}}
\newcommand{\ben}{\begin{eqnarray*}}
  \newcommand{\een}{\end{eqnarray*}}
\newcommand{\bea}{\begin{eqnarray}}
  \newcommand{\eea}{\end{eqnarray}}
\newcommand{\bdm}{\begin{displaymath}}
  \newcommand{\edm}{\end{displaymath}}
\newcommand{\ba}{\begin{align}}
  \newcommand{\ea}{\end{align}}

\newcommand{\E}{\text{e}}
\newcommand{\I}{\text{i}}
\newcommand{\D}{\text{d}}

\usepackage{tensor}

\begin{document}

\title{Quantum Theory of the Lema\^{\i}tre Model for\\
  Gravitational Collapse}

\titlerunning{}        

\author{Claus Kiefer\orcidlink{0000-0001-6163-9519}        \and Hamid Mohaddes
}

\institute{C. Kiefer \at
              Institute for Theoretical Physics,
              University of Cologne\\
              Z\"ulpicher Stra\ss e 77,
              50937 K\"oln, Germany\\
              \email{kiefer@thp.uni-koeln.de}  
           \and
H. Mohaddes \at
School of Mathematical Sciences,
              University of Nottingham\\
              Nottingham NG7 2RD, United Kingdom
 \\ \email{Hamid.mohaddes@nottingham.ac.uk} \\ 
}

\date{Received: date / Accepted: date}

\maketitle

\begin{abstract}

We investigate the quantum fate of the classical singularities that
occur by gravitational collapse of a dust cloud. For this purpose, we
address the quantization of a model first proposed by Georges
Lema\^{\i}tre in 1933. We find that the singularities can generically
be avoided. This is a consequence of unitary evolution in the quantum
theory, whereby the quantum dust cloud collapses, bounces at a minimal
radius and re-expands. 
  
\end{abstract}


\vskip 3mm

\noindent {\em Invited contribution to the Proceedings of the
  Lema\^{\i}tre Conference 2024 at Specola Vaticana}

\section{Introduction}

One of George Lema\^{\i}tre's most important papers is {\em L'univers
  en expansion} published in 1933 \cite{GL33}. As Andrzej Krasi\'nski
emphasizes in his editorial notes
following the English translation of \cite{GL33}, this paper plays a
pioneering role for various reasons \cite{AK97}.

Perhaps the main reason is that the paper contains
a derivation of a spherically-symmetric solution of inhomogeneous dust
from Einstein's field equations. In this way, it generalizes
Schwarzschild's solutions for both vacuum and homogeneous static
dust. This solution plays a major role in our contribution.
But there are at least three other important developments initiated
in \cite{GL33}. First, Lema\^{\i}tre suggests a possible
mechanism to describe the formation of clusters of galaxies (called
{\em n\'ebuleuses} there). Second, the paper contains a proof that
the Schwarzschild horizon at $r=2GM/c^2$ is only a
    coordinate singularity. And third, it presents an introduction to
    the concept of Misner--Sharp mass thirty years before Misner
      and Sharp published their work.   

Lema\^{\i}tre worked on his paper during his visit to
the United States in 1932--33, in particular during his stay at the {\em Caltech}
in Pasdadena from November 1932 to January 1933, where he also met
Albert Einstein (Fig.~1).

\begin{figure}[h]
  \centering
  \includegraphics[width=0.8\textwidth]
  {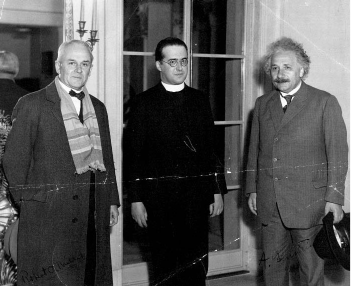} 
 \caption{Robert Millikan, Georges Lema\^{\i}tre, and Albert Einstein
   at California Institute of Technology, January~1933. Figure credit:
   Wikimedia Commons} 
\end{figure}

The most intense interaction concering the topic of \cite{GL33} during
his stay at {\em Caltech} was with Richard Tolman, who himself wrote a
paper on this topic in 1934. As discussed in \cite{AK97}, although
Tolman gives explicit credit to  Lema\^{\i}tre, the inhomogeneous dust
solution became known as ``Tolman model'' or ``Tolman--Bondi model''
after a paper by Hermann Bondi on this topic had appeared in 1947.
We shall follow here the more recent practice of calling this solution
the Lema\^{\i}tre--Tolman--Bondi solution or LTB solution. For details
on biographic aspects, we refer to \cite{Lambert}.

The LTB model has since been used extensively in classical
relativity, in particular in addressing questions of structure formation; see,
for example, \cite{krasinski97} for details of the classical
theory. Here, instead, we use 
this model as a starting point for quantization, in order to get
insights into how the classical picture of gravitational collapse may be
modified in the quantum theory. This allows to address questions such
as what is the fate of the black-hole singularity or what is the role
of white holes.

Our paper is organized as follows. Before turning to our main topic,
we shall briefly address the simpler situation of a
thin null dust shell. Then we discuss the LTB model at the classical
and at the quantum model. A major issue is to investigate the
quantum version of the classical collapse scenario. We shall, in fact,
see that the classical singularity can be avoided and that the
initial collapse of a wave packet mimicking a shell in the dust cloud
will be followed by its re-expansion. We then discuss the simpler case
of the Oppenheimer--Snyder (OS) scenario, which is obtained from the
LTB model in the limit of constant density \cite{OS39}. This
simplification allows the derivation of more explicit details.
We shall end with a brief summary and
reflections about future developments.

\section{Thin null dust shell}

Before starting our discussion for the dust cloud in the LTB model, we shall
briefly address the simpler case of a single self-gravitating {\em
  dust shell}. It has 
turned out that the case of a null dust shell is especially suitable
for our purpose; see \cite{OUP3,Hajicek:2001yd,Hajicek-review} and the references
therein for details.

Classically, the shell either collapses to a black-hole singularity or
expands from a white-hole singularity. A consistent quantum version
can be obtained by the method of reduced quantization. The shell can
there be represented by a quantum state $\Psi_{\kappa\lambda}(t,r)$,
where $t$ is the asymptotic (Killing) time, $r$ the shell radius and
$\kappa$ (positive integer) and $\lambda$ (positive number with
dimension of a length) are two parameters characterizing the wave
packet that is the quantum version of the shell. At $t=0$, we choose
the following family of wave packets in momentum space:
\begin{equation}
  \label{packet-shell}
  \psi_{\kappa\lambda}(p) :=
   \frac{(2\lambda)^{\kappa+1/2}}{\sqrt{(2\kappa)!}}
  p^{\kappa+1/2}\E^{-\lambda p}.
\end{equation}
By an appropriate choice of $\kappa$ and $\lambda$, a narrow wave
packet can be constructed. After an integral transform from the $p$- to
the $r$-transformation, one can find an exact solution to the
Schr\"odinger equation. It reads
\be
\label{psi-shell}
  \Psi_{\kappa\lambda}(t,r) = \frac{1}{\sqrt{2\pi}}
  \frac{\kappa!(2\lambda)^{\kappa+1/2}}{\sqrt{(2\kappa)!}}
  \left[\frac{\I}{(\lambda +\I t +\I r)^{\kappa+1}}
   - \frac{\I}{(\lambda +\I t-\I r)^{\kappa+1}}\right]. 
 \ee
 An important property of this solution is that the wave function
 vanishes at the position $r=0$ of the classical singularity,
 \be
  \lim_{r\rightarrow 0}\Psi_{\kappa\lambda}(t,r)= 0.
\ee
This means that the probability of finding the shell at
vanishing radius is zero! In this sense, the singularity is avoided
in the quantum theory. The quantum shell bounces and re-expands, and
no event horizon forms.

From \eqref{psi-shell}, we can find the expectation value
\bdm
\langle R_0\rangle_{\kappa\lambda}:= 2G\langle E\rangle_{\kappa\lambda}
=(2\kappa +1)\frac{l_{\rm P}^2}{\lambda},
\edm
of the shell radius and its variance,
\bdm
\Delta(R_0)_{\kappa\lambda}:=2G\Delta E_{\kappa\lambda}=
\sqrt{2\kappa +1}  \frac{l_{\rm P}^2}{\lambda},
\edm
where $l_{\rm P}=\sqrt{G\hbar}$ is the Planck length ($c=1$ here).
It thus turns out that the wave packet can be squeezed
below its Schwarzschild radius if its energy is greater
than the Planck energy---a genuine quantum effect!

In a sense, one can describe this scenario as a ``superposition of
black and white hole'': the quantum solution contains information
about the classical black hole {\em as well as} the classical white
hole solution. The two together enable a singularity-free quantum
state. Similar features were found later in loop quantum gravity
\cite{HR15} and in quantum cosmology \cite{KZ95}. 

This is an interesting result, but it emerges from a simple model that
may not reflect the situation in the real world. The model can only be in
accordance with observations if the timescale between collapse and
re-expansion is sufficiently long, that is, if it is at least comparable with
the age of the Universe (because we have so far no evidence for a
reversal of collapsing stars). It is certainly imaginable that
gravitational time delay is sufficiently long to guarantee this
consistency. But explicit calculations are not simple, mainly due to
the problem of {\em defining} an appropriate time delay between
comoving and stationary observer (naively, for a stationary observer
there is an infinite time delay). In fact, 
different results have appeared in the literature; see, for example, the review
\cite{Malafarina:2017csn}. Timescales $\propto M^3$ or $M^2$, where $M$ is the
initial mass, may be sufficient for this purpose, but the issue is not
settled. This question will also be of relevance for the LTB model.

Singularity avoidance in this model is reflected by the fact that the
quantum state vanishes at the place of the classical singularity. This
is here a consequence of the self-adjoint nature of the reduced
Hamiltonian, which leads to a unitary time evolution. Unitarity
prevents the wave packet from disappearing in a singularity -- the packet must
always be present somewhere. From this point of view, it is not
surprising that the packet collapses, bounce, and re-expands. This is
what we shall also find in the LTB model.

In quantum cosmology, one often imposes from the outset $\Psi=0$ at
places of classical singularities. This was already suggested by Bryce
DeWitt in 1967 \cite{Witt67} and is called DeWitt criterion. While
this is related in spirit to the vanishing of quantum states in our
case, it is not equivalent because in quantum cosmology there is no
asymptotic time and no unitarity -- these concepts emerge there only in a
semiclassical approximation \cite{OUP3}. 

\section{LTB model}
\label{ch:chapter_2}

In this section, we shall briefly introduce the classical LTB
model. It describes a spherically-symmetric
    solution of Einstein's equations with non-rotating dust of energy (mass)
    density $\epsilon (\rho) $ as its source (where $\rho$ denotes the
    radial coordinate), see \eqref{LTB1} below. For constant energy density, we
    arrive at the Oppenheimer--Snyder (OS) scenario, which provided
    the first example of a solution describing the gravitational
    collapse to what was later called a black hole \cite{OS39}.
    In the case of dust (no pressure or viscosity), we can
    interpret the cloud as consisting of infinitely many {\em independent shells}.
    This will be mandatory for
    developing the formalism of quantization.

    The line element of the LTB solution can be written in the form
    \begin{align}
\label{LTB1}
    \D s^2&=-c^2 \D\tau^2+\frac{R^\prime}{1+2f}\D\rho^2+R^2\D\Omega^2,\\
    \label{LTB2}
    \frac{8\pi G}{c^2}\epsilon&=\frac{F^\prime}{R^2R^\prime},\\
    \label{LTB3}
    \frac{\Dot{R}^2}{c^2}&=\frac{F}{R}+2f,
\end{align} 
where $\tau$ is the dust proper time and a prime denotes a derivative
with respect to the radial variable $\rho$ 
        that labels the dust shells comprising the dust
        cloud; $F(\rho)$ is twice the active gravitational mass (the
        Misner--Sharp mass $M$) inside the shell with label
        $\rho$.\footnote{Inserting the gravitational constant $G$ and
          the speed of light $c$, we have $F=2GM/c^2$.} We
        restrict ourselves to vanishing cosmological constant. The
        variable $f$ is a measure of the curvature of the subspaces
        with constant time; below we restrict ourselves to the
        marginally bound (flat) case $f=0$.
        
An important quantity is $R(\tau,\rho)$, which is the curvature radius of
the shell labelled by $\rho$ at time $\tau$. It describes how the
shell collapses or expands. A central singularity forms at $R=0$. There
are also singularities from shell crossings happening when
two dust shells occupy the same radius. In the simplified setting
below, these do not occur.
        
Let us now address the quantization of this model and its consequences
for the collapse situation. One possibility is to apply canonical
(Wheeler--DeWitt) quantization. This has led to interesting results
concerning Hawking radiation and black-hole entropy, but did not allow
finding exact quantum states \cite{OUP3}. Exact results can be obtained
if one restricts oneself to self-gravitating dust clouds without
introducing additional quantum fields, that is, without the
possibility to implement Hawking radiation. This is what we shall
review here. 
In this, we shall follow \cite{KS19} to which
we refer the reader for more details.

As for the case of the thin shell above, we employ here the
method of {\em reduced quantization}.
We assume that 
the infinitely many different shells comprising the cloud decouple, so we can focus on a
single shell, here: the outermost shell. This simplifies the
calculations drastically. An early analysis along these lines was presented
by Fernando Lund \cite{Lund73}, who obtained qualitative results about singularity
avoidance without calculating exact quantum wave packets. 
In order to derive exact solutions, we start with the
Hamiltonian for the outermost shell 
(with radius $R_o$) given by
\be
\label{Hamiltonian-o}
H=-\frac{P_o^2}{2R_o},
\ee
which is the negative of the ADM energy.
($P_o$ is the momentum conjugate to $R_o$.) As mentioned above,
restriction is made here to the 
marginally bound case of the LTB solution ($f=1$ in \eqref{LTB1}).

The fact that the Hamiltonian \eqref{Hamiltonian-o} is negative might
seem surprising. But this is not unusual for gravitational systems
because it reflects the attractivity of gravity \cite{GK94}. The
physical (ADM) energy is positive.

As in the case of the collapsing shell, we seek for a unitary
evolution (here with respect to the dust proper time
$\tau$). In the Schr\"odinger representation, we have
\bdm
P_o\to\hat{P}_o=-\I\hbar\frac{d}{d R_o}.
\edm
The operator $\hat{R}_o$ acts by multiplication. (In the following,
we shall suppress the subscript $o$.) The Hamilton operator is then
given by
\bdm
\hat{H}=\frac{\hbar^2}{2}~R^{-1+a+b}\frac{d}{d
  R}R^{-a}\frac{d}{d R}R^{-b},
\edm
where $a$ and $b$ encode factor ordering ambiguities. The
Schr\"odinger equation reads
\be
\label{Schroedinger}
\I\hbar\frac{\partial\Psi(R,\tau)}{\partial\tau}=\hat{H}\Psi(R,\tau).
\ee
  We impose square-integrability on wave functions and let
them evolve unitarily according to a self-adjoint Hamiltonian.
This corresponds to enforcing  probability conservation in dust proper
time. The dynamics of the resulting wave packets will be presented in the next
section.

\section{Singularity avoidance}

To solve the Schr\"odinger equation,
 an initial quantum state (e.g. at $\tau=0$) must be specified. In
 order to find the quantum version of the classical collapse, we
 choose a narrow initial wave packet that mimicks the classical shell.
A detailed investigation shows that for a wide class of wave packets,
the probability for the outermost 
   dust shell to be in the classically singular configuration $R=0$ is
   {\em zero} \cite{KS19}. To give one example for an exact solution
   of \eqref{Schroedinger}:
   \begin{multline}
     \label{wave-packet}
\Psi(R,\tau)=\sqrt{3}\left(\!\frac{\sqrt{2}}{3}\right)^{\frac{1}{3}\left|
    1+a\right|+1}\frac{\Gamma\!\left( \frac{1}{6}\left|
      1+a\right|+\frac{\kappa}{2}+1\right)
}{\sqrt{\Gamma(\kappa+1)}\Gamma\!\left( \frac{1}{3}\left|
      1+a\right|+1\right)}~R^{\frac{1}{2}(1+a+|1+a|+2b)}\\ 
\quad\times\frac{\lambda^{\frac{1}{2}(\kappa+1)}}{(\frac{\lambda}{2}-\I\tau)^{
    \frac{1}{6}\left| 1+a\right|+\frac{\kappa}{2}+1}}~_1F_1\!\left(
  \tfrac{1}{6}\left| 1+a\right|+\tfrac{\kappa}{2}+1;
  \tfrac{1}{3}\left| 1+a\right|+1;-\frac{2
    R^3}{9(\frac{\lambda}{2}-\I\tau)}\right),
\end{multline}
where $_1F_1$ denotes Kummer's confluent hypergeometric function.
The parameters $\kappa$ and $\lambda$ have the same meaning as in the
thin-shell case above; see Eq.~\eqref{packet-shell}. 
This solution is plotted in Fig.~2, where it is also compared with the
classical trajectory.

\begin{figure}[h!]
\centering
    \includegraphics[width=95mm]{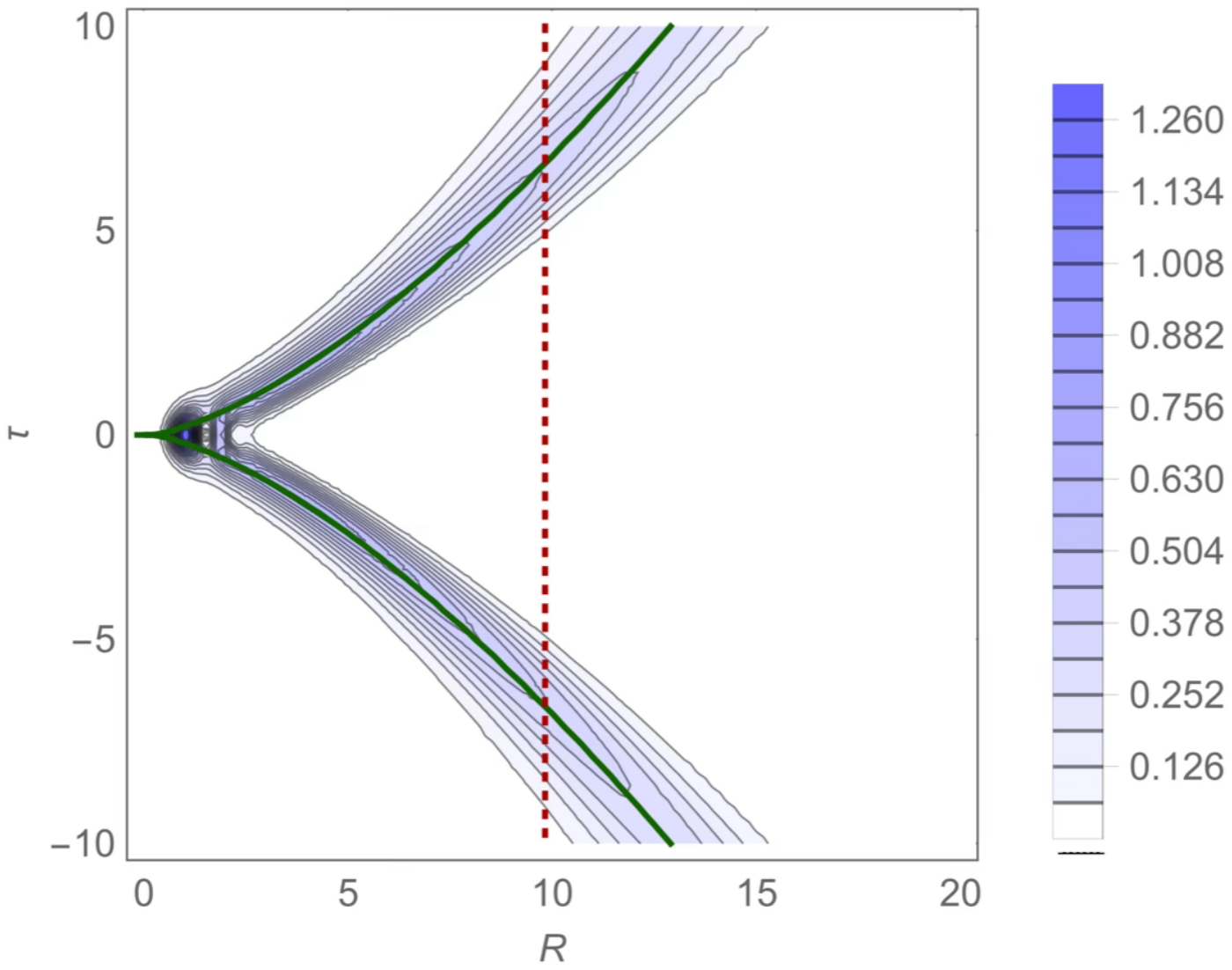}
    \caption{Probability amplitude for $R$ as given by
        $R^{1-a-2b}~|\Psi(R,\tau)|^2$, see \eqref{wave-packet},
        compared to the classical
          trajectories (full green line) and the exterior
          apparent horizon (dotted red
          line), with $a=2$ and $b=1$, and $\lambda=2.2$,
          $\kappa=0.96$.  Reproduced from \cite{KS19} with kind
        permission by the American Physical Society.}
    \label{fig:2}
  \end{figure}

We recognize that the packet first follows the
infalling classical trajectory up to some minimal radius $R$ and then
makes a transition to the outgoing classical trajectory.
The behaviour of the wave packet thus resembles the behaviour for the thin
shell described above. The packet first collapses, enters the apparent
horizon, but then bounces and re-epands. This is, again, a consequence
of the unitary evolution which follows from having a self-adjoint
reduced Hamiltonian. In this sense it is different from imposing the
DeWitt criterion of vanishing wave function from the outset.

As for the thin shell above, a crucial issue is the lifetime of the
bouncing cloud (or ``Planck star''), that is, the elapsed time between
collapse and re-expansion. For this purpose, two observers are
introduced, one at a fixed physical radius outside the object
(stationary observer), one comoving with the cloud. These observers
meet twice -- first during collapse and second during
re-expansion. Applying the method for calculating the lifetime in a
similar situations for spinfoams \cite{CA24}, one finds here a
lifetime $\propto M^3$, which would be long enough to be in accordance
with the observational non-evidence for such objects. In fact, a
lifetime $\propto M^3$ is distinguished because it coincides with the
lifetime of black holes due to emission of Hawking radiation and also
with the spreading time for wave packets in models of stationary quantum black
holes \cite{KL99}. 
  
\section{Oppenheimer--Snyder model}

An important special case of the LTB model is the Oppenheimer--Snyder
(OS) model \cite{OS39}. Here, the collapsing dust cloud is
homogeneous. The geometry is then described in the exterior by the
Schwarzschild metric and in the interior by the
Friedmann-Lema\^{\i}tre-Robertson-Walker (FLRW) metric. As in
cosmology, one can distinguish between flat, closed, 
and open FLRW metric. 

Dirac quantization of this model for the flat FLRW case (with
curvature parameter $k=0$) is attempted
in \cite{Tim20}. In 
contrast to the above discussion, the whole cloud is treated, not just
the outermost shell as a representative of the cloud. Using
Brown-Kucha\v{r} dust for matter and employing the standard Kucha\v{r}
decomposition in the canonical formulation \cite{OUP3}, one arrives at
a form of the Hamiltonian constraint that is of a multivalued nature
and looks too complicated for a direct application of Dirac
quantization. Nevertheless, some preliminary results can be obtained
\cite{Tim20}.

A more promising method is affine coherent state
quantization (ACSQ) \cite{PS20}. Coherent state quantization relies on the
identification of the classical phase space with a Lie group.
One can then consider a unitary irreducible representation of the
group on a suitable Hilbert space, letting it act on a fixed state.
This
allows constructing a family of coherent states in the quantum
theory. The affine group comes into play because phase space here
corresponds to a
half line. If the phase-space function is semi-bounded, the resulting
operator after quantization is self-adjoint.
A brief self-contained introduction to affine quantization can be found in
\cite{PS20}.

Using again the flat FLRW case for the OS dust cloud, the authors of
\cite{PS20} find that both the comoving and the stationary observers
see a bounce: the OS cloud collapses to a minimal radius outside the
photon sphere and then re-expands. Because of this large minimal radius,
one cannot even speak of a black hole. The lifetime seen by the comoving
observer is again proportional to $M$, but it was not possible to
define a suitable lifetime for the stationary observer. The question
of whether this scenario reflects features of the real world thus
remains unanswered. 

The closed and open OS models, also called nonmarginal models
(curvature parameters $k=1$ and $k=-1$), 
are discussed in \cite{KM23}. There again
the method of ACSQ is used. The discussion is more involved than in
the flat case, but one finds again that under some conditions
there is a bounce of the collapsing cloud, as seen both from the viewpoint of
the comoving and of the stationary observers.  
An interesting special scenario for the closed case (particular choice
of parameters) is shown in Fig.~3. 
    
\begin{figure}[h]
  \includegraphics[width=0.6\textwidth]
  {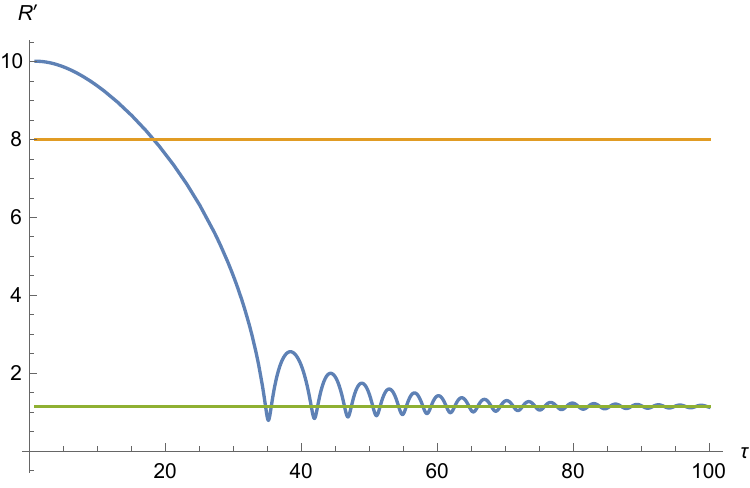} 
  \caption{Graph in the $R-\tau$ space. The orange line
      represents the Schwarzschild radius $R=2GM/c^2$, the green line the 
      location where equilibrium is reached after oscillations, where
      $\dot{R}=\ddot{R}=0$. Reproduced from \cite{KM23} with kind
        permission by the American Physical Society.}
\end{figure}

A comoving observer obtains the following picture.
Inside the horizon, the curve $R(\tau)$ exhibits oscillations. This is
different from the flat case and means that the cloud collapses,
experiences a bounce and oscillates until it reaches an equilibrium.
This looks as if from the outside this situation could not be distinguished from 
a Schwarzschild black hole. Unfortunately, it turns out that a horizon
never forms from the viewpoint of the stationary observer and that
therefore the object does 
not resemble a black hole. This apparent conflict remains an open
issue and is subject for further discussion.

\section{Summary and outlook}

We hope we have convinced the reader that Lema\^{\i}tre's model from
1933 \cite{GL33} is not only well suitable for problems in classical
cosmology, but also for addressing fundamental issues in quantum
gravity. As we have reviewed here, one can construct quantum models
for gravitational collapse which are {\em singularity-free}, that is,
both the classical black-hole as well as the classical white-hole
singularities can be avoided. There is, in fact, a unitary evolution for
the quantum state from a collapsing wave packet to a bounce and a
re-expanding packet. In a sense, the quantum theory knows of both
black holes and white holes, even if classically white holes are
irrelevant. If generally true, this would solve the
cosmic-censorship problem
because singularities are fundamentally absent.

In spite of these promising results, there remain open
problems. First, these results were found for spherically-symmetric
systems only. Real black holes are described by the Kerr metric, but
an analysis similar to the one presented here seems, at present,
impossible for the rotating case. Second, there remains the problem of how
the lifetime of these collapsing and re-expanding wave packets can be
properly defined for stationary observers. There will be no conflict with
observation only if the time delay between collapse and expansion is
at least of the order of the age of our Universe. And third, these
models should be extended by taking into account appropriate quantum
matter fields so that the issues of Hawking radiation and information loss
can be discussed at an exact quantum gravity level. Perhaps the
groundbreaking work of Georges Lema\^{\i}tre will continue to guide us
towards that goal.


\begin{acknowledgements}
C.K. thanks the organizers of the conference Lema\^{\i}tre~2024 for
inviting him to a wonderful and inspiring meeting.
\end{acknowledgements}

\vskip 5mm

\noindent {\bf Data Availability Statement}: Data sharing not
applicable--no new data generated.



\begin{thebibliography}{99}

 \bibitem{GL33}  Lema\^{\i}tre, G.: 
           Ann. Soc. Sci. de Bruxelles
           {\bf A 53}, 51--85 (1933); for an English translation, see:
             Gen. Relativ. Gravit. {\bf 29}, 641--680
             (1997)

 \bibitem{AK97} Krasi\'nski, A.: Gen. Relativ. Gravit. {\bf 29},
   637--640 (1997)
   
 \bibitem{Lambert} Lambert, D.: Un atome d'univers, \'Editions
   Lessius, Bruxelles (2000)
   
    \bibitem{krasinski97}
      Krasi\'nski, A.:
       Inhomogeneous Cosmological Models, Cambridge University
       Press (1997)

   \bibitem{OS39} Oppenheimer, J. R. and Snyder, H.: Phys. Rev. {\bf
       56}, 455--459 (1939)
       
  \bibitem{OUP3} Kiefer, C.: Quantum Gravity,
    International Series of Monographs on Physics, 3rd edn.,
    Oxford University Press, Oxford (2012).
    
	\bibitem{Hajicek:2001yd} H{\'a}j{\'i}\v{c}ek, P. and
          Kiefer, C.: Int. J. Mod. Phys. D {\bf 10}, 775--780 (2001)

        \bibitem{Hajicek-review}  H{\'a}j{\'i}\v{c}ek, P.:
          Lect. Notes Phys. {\bf 631}, 255--299 (2003).

        \bibitem{HR15} Haggard, H. M. and Rovelli, C.:  Phys. Rev. D
          {\bf 92}, 104020 (2015)

           \bibitem{KZ95} Kiefer, C. and Zeh, H.D.: Phys. Rev. D {\bf
               51}, 4145--4153 (1995).
    
          \bibitem{Malafarina:2017csn} Malafarina, D.: Universe {\bf
              3}, 48 (2017)          

          \bibitem{Witt67} DeWitt, B. S.:  Phys. Rev. {\bf 160},
            1113--1148 (1967)
            
   \bibitem{KS19} Kiefer, C. and Schmitz, T.: Phys. Rev. D {\bf
              99}, 126010 (2019)
            
          \bibitem{Lund73} Lund, F.: Phys. Rev. D {\bf 8}, 3253--3259
            (1973)

       \bibitem{GK94}  Giulini, D. and Kiefer, C.: Phys. Lett. A {\bf
           193}, 21--24 (1994)

         \bibitem{CA24} Christodoulou, M. and D'Ambrosio, F.:
           Class. Quantum Grav. {\bf 41}, 195030 (2024)

           \bibitem{KL99} Kiefer, C. and Louko, J.:
             Ann. Phys. (Berlin) {\bf 8}, 67--81 (1999)
             
           \bibitem{Tim20} Schmitz, T.:  Phys. Rev. D {\bf 101}, 026016 (2020)         

            \bibitem{PS20} Piechocki, W. and Schmitz. T.:
              Phys. Rev. D {\bf 102}, 046004 (2020)

            \bibitem{KM23} Kiefer, C. and Mohaddes, H.: Phys. Rev. D {\bf
              107}, 126006 (2023)                     


\end{thebibliography}
\end{document}